# Partial *D*-operators for the generalized IBP reduction

A.A. Radionov and F.V. Tkachov

Institute for Nuclear Research of Russian Academy of Sciences, Moscow 117312

**Abstract.** Empirical evidence reveals existence of partial *D*-operators for the generalized IBP (BT) reduction algorithms that are, counterintuitively, much simpler and much easier to find than the complete *D*-operators from the foundational Bernstein theorem, allowing one to construct first true two-loop examples of generalized IBP identities.

**1 Introduction.** The IBP reduction algorithms [1] are a standard tool of multiloop calculations. They exploit differential identities based on algebraic structure of multiloop integrands to express complex integrals in terms of simpler ones. In ref. [2], a rather general version was suggested, based on a fundamental theorem due to Bernstein [3].

These generalized IBP identities are formulated for integrals in parametric representation and aim to make singular integrands amenable to numerical evaluation. However, finding the corresponding differential operators proved to be extremely hard. In practical calculations, only the explicit solution for the one-loop case found in [2] was used within the so-called minimal BT-approach [4], and some non-trivial but partial results related to 2-loop integrals were published by the GRACE collaboration [5]. In ref. [6], we announced an array of findings that allowed us to obtain first true two-loop examples of generalized IBP identities. In this Letter we focus on the theoretical part of the findings, namely, the use and properties of the so-called *partial D-operators*. As regards software, we only mention here that for all calculations, we employed a custom computer algebra engine Gulo [7] built on the foundation of the Oberon technologies (for a review see [8]) and significantly improving upon an earlier version that was successfully used in demanding calculations [9].

**2 Setup**. Standard dimensionally regularized multiloop integrals with two or more loops in the Feynman-Chisholm representation [10] have the following general form:

$$\int_S dx \sum_{n_0,n_1} P_0^{\mu_0}(x) P_1^{\mu_1}(x) R_{n_0,n_1}(x) \tag{1}$$

where $x$ is a vector of real variables, the integration domain $S$ is a simplex $S = \{x \mid \forall i : x_i > 0; \Sigma_i x_i < 1\}$, $P_i$ (known as Symanzik polynomials, see e.g. [11]) and $R_{n_0,n_1}$ are polynomials of $x$, the summation over integer $n_0, n_1$ runs over a finite region, $\mu_i = -n_i + c_i(D-4)$ where $c_i$ are small rational numbers. It is practically sufficient to regard $\mu_i$ as independent symbolic variables. One of the $P_i$ depends only on the topology of the corresponding Feynman diagram and has degree $l$ (the number of loops), whereas the other one has degree $l+1$ and its coefficients depend linearly on the kinematical parameters (masses and momenta) of the physical process. The summation over $n_0, n_1$ and the polynomials $R_{n_0,n_1}$ (that also depend on kinematical parameters) is a feature of non-scalar particle models. This complication does not affect the principle of use of the differential identities considered below.

The essential difficulty with numerical evaluation of such integrals is that $P_i$ may have singularities within and/or on the boundaries of $S$, whereas the sum may involve $n_0, n_1$ large enough to render the integrand non-integrable even if the integral is meaningful via analytical continuation in $\mu_i$.

Bernstein proved [3] that for any polynomial of several variables $P(x_0, x_1, \ldots)$ there exists a finite-order differential operator $D(\partial_0, \partial_1, \ldots)$, where $\partial_i$ is partial derivative with respect to $x_i$, such that:

$$D(\partial_0, \partial_1, \ldots) P^\mu(x_0, x_1, \ldots) = b(\mu) P^{\mu-1}(x_0, x_1, \ldots), \quad b(\mu) \neq 0 \tag{2}$$

One can say that the operator $D$ decreases the power of $P$ by one. The coefficients of $D$ are polynomials of $x_i$, $\mu$. The essence of Bernstein's proof is that if one takes $D$ of increasing degrees with unknown coefficients, then, after reducing the identity to a homogeneous linear system, the number of unknowns grows faster than the number of equations, whereas the differentiations ensure a mixing of monomials of different original degrees with respect to $x$, so that with sufficiently large degrees of $D$ with respect to $\partial$ the system saturates and begins to have non-zero solutions. This theorem holds irrespective of algebraic nature of the coefficients of $P$ — numbers or polynomials of other symbolic parameters.

For applications to integrals of the form (1) one needs a (trivial) generalization of eq. (2) to the case of several (usually two) polynomials $P_i$:

$$D(\partial_0, \partial_1, \ldots) \prod_i P_i^{\mu_i}(x_0, x_1, \ldots) = b(\mu) \prod_i P_i^{\mu_i - 1}(x_0, x_1, \ldots), \quad b(\mu) \neq 0 \qquad (3)$$

The generalized IBP reduction consists in applying the formula several times, each time replacing the product of two polynomials raised to complex powers (see (1)) by the left-hand side of eq. (3) and getting rid of derivatives via integration by parts. As a result, the powers of the dangerous polynomials are raised to values that are safe for numerical integration, whereas the boundary terms that are generated by the integration by parts would have the same form as (1) with a lesser dimensionality, i.e. a lesser number of $x_i$.

For many interesting applications within perturbative QFT, it is necessary to find ways to construct the operator $D$ for a given pair of polynomials $P$. Since there is only an existence theorem of unspecific nature for $D$, the basic way to search for a solution is via the method of undetermined coefficients: first one makes an assumption about the degrees of $D$ with respect to $\partial$ and $x$, then $D$ is written out with undetermined coefficients, finally eq. (2) is reduced to a system of homogeneous linear equations for the undetermined coefficients. Bernstein's result guarantees that, starting from sufficiently large degrees of $D$, the system will have non-zero solutions with $b(\mu) \neq 0$. This basic method can be optimized in abstract mathematical (cf. [5]), and/or specific algorithmic (cf. [6]) fashions.

For one-loop integrals, the topological polynomial degenerates into 1, whereas the kinematical one is quadratic in $x$, and an explicit solution for $D$ can be written out [2]. This special case proved indeed useful for some numerical 2-loop calculations [4].

The GRACE collaboration in ref. [5] presented first non-trivial examples of *D*-operators related to two-loop integrals but only with one, kinematical polynomial (for the vertex diagram considered in [5], it has third degree and depends on five independent $x_i$). Ref. [5] used an abstractly formulated mathematical algorithm and two general-purpose computer algebra systems to obtain a solution for $D$. After cancelling a common numerical factor their solution fits on three lines (see [12] for the simplified expression).

Unfortunately, the inclusion of the second, topological polynomial into the problem (it has degree 2 with respect to $x$) complicates things rather significantly. As a rule of thumb, the complexity of construction of $D$ for two polynomials is to be compared to that for a single polynomial equal to the product of the two. Since the resulting task proves to be extremely cumbersome, it is essential to eliminate the implicit prejudicial assumptions from the problem formulation. We thus consider (following in the footsteps of [5]) only the case of $P$ with numerical (rational) coefficients, which corresponds to setting the kinematic parameters to specific values (this assumes that amplitudes computed for various such sets of values are to be used for a subsequent interpolation to build, typically, a Monte Carlo generator). The resulting analytical answers are not in general expected to be suitable for human viewing but to be fed directly into a computer algebra system to perform the IBP reduction.

**3 Partial *D*-operators.** In [6], we discussed a calculational scheme that allows one to construct the operators $D$ rather efficiently (under one second on a low-end laptop) at least for the simpler of the non-degenerate 2-loop cases. Along with purely programmatic optimizations, the method of [6] employs a significant theoretical finding that is independent of implementation details, which we now proceed to describe.

The first point to notice is a significant asymmetry between the integer parts $n$ of the powers $\mu$ in loop integrals (1) in models with non-scalar particles. So, it may be advantageous to raise the power of only one polynomial at one step of the reduction process. One is thus led to consider what we call partial *D*-operators that affect the power of only one factor:

$$D_0(x,\mu_0,\mu_1,\partial) P^{\mu_0} P^{\mu_1} = b_0(\mu_0,\mu_1) P^{\mu_0-1} P^{\mu_1}, \quad b_0(\mu_0,\mu_1) \neq 0 \tag{4}$$

$$D_1(x,\mu_0,\mu_1,\partial) P^{\mu_0} P^{\mu_1} = b_1(\mu_0,\mu_1) P^{\mu_0} P^{\mu_1-1}, \quad b_1(\mu_0,\mu_1) \neq 0 \tag{5}$$

Such a pair of partial *D*-operators can be composed into a complete one that satisfies (3):

$$\begin{aligned} D(x,\mu_0,\mu_1,\partial) &= D_1(x,\mu_0-1,\mu_1,\partial) D_0(x,\mu_0,\mu_1,\partial) \\ b(\mu_0,\mu_1) &= b_0(\mu_0,\mu_1) b_1(\mu_0-1,\mu_1) \end{aligned} \tag{6}$$

There are three a priori observations to be made here:

1. Partial *D*-operators always exist. For instance, to obtain (4), wherein only the power of $P_0$ is raised, it is sufficient to multiply both sides of (3) by $P_1$. But the partial $D_0$ thus obtained is rather more complex than the complete $D$.

2. If one tries to find a partial operator directly via undetermined coefficients, then one finds that the resulting system of equations does not look simpler than the one for the complete operator, so it is not a priori evident that the partial *D*-operators might be easier to find, nor that they might be simpler than the complete ones.

3. One cannot a priori expect that the *simplest* complete operators (i.e. the ones with lowest degrees in $x$ and $\partial$) are representable as the above composition of partial ones, eq. (6) — and if not, then the partial *D*-operators cannot fully replace the complete ones.

The above arguments prevent one from expecting the partial *D*-operators to be useful. This expectation, however, was proved wrong by experiments made possible by the efficient software employed. Our findings (derived from a variety of examples) are as follows:

i) In regard of overall complexity, simplest (i.e. ones with lowest degrees with respect to $x$ and $\partial$) partial *D*-operators look like factors of simplest complete *D*-operators, i.e. their composition (6) yields a complete *D*-operator of the same complexity as ones constructed directly. (An exception to this rule is the boundary case of two linear polynomials where the simplest partial and complete *D*-operators have degree 1 with respect to derivatives.) Remember that there is always a family of solutions, and one cannot expect any complete *D*-operator to be thus factorizable.

ii) The simplest partial operators are constructed correspondingly faster than the complete ones — i.e. very significantly faster due to the non-linearity of the construction time as a function of the degrees of *D*-operators.

iii) The relative complexity and construction time of the partial *D*-operators follow the relative complexity of the corresponding polynomials *P*.

We have so far found no counterexamples to these findings. Given the generic nature of the mechanism behind the Bernstein theorem, it is reasonable to adopt these findings as a guiding principle for further explorations.

iv) A useful observation that admits a direct proof is as follows:

**Theorem**. If (non-constant) $P_i$ and $\prod_{j \neq i} P_j$ do not have a non-trivial common divisor then $b_i$ has $\mu_i$ as a factor.

This holds as a rule for Symanzik polynomials with rare and marginal exceptions, e.g. for $k^2 = 0$ in the example below where $P_0$ then divides $P_1$ and the solution simplifies greatly.

**4 Example of a non-trivial partial 2-loop *D*-operator**. Whereas multi-page analytical answers for multiloop integrals are pretty common, the *D*-operators — although analytical — are certainly not meant for human viewing. Therefore, we only present a simple example in order to give one a general impression, referring interested readers to a dedicated web page [12] where further, longer examples are posted. Consider the simplest 2-loop self-energy of the model $g\varphi^4$,

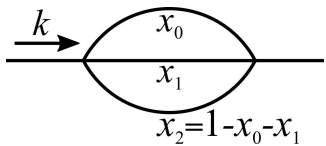

The corresponding two Symanzik polynomials for $m_0^2 = 1$, $m_1^2 = 0$, $m_2^2 = 0$, $k^2 = 2$ are as follows ($x_2 \equiv 1 - x_0 - x_1$):

$$P_0 = x_0 x_1 + x_0 x_2 + x_1 x_2 = x_0 + x_1 - x_0^2 - x_1^2 - x_0 x_1 \tag{7}$$

$$P_1 = \left(m_0^2 x_0 + m_1^2 x_1 + m_2^2 x_2\right) P_0 - k^2 x_0 x_1 x_2 = x_0^2 - x_0 x_1 - x_0^3 + x_0^2 x_1 + x_0 x_1^2 \tag{8}$$

The partial operator $D_0$ and the corresponding Bernstein polynomial $b_0$ are found using the software described in [6] within a few seconds on a low-end laptop computer. The results are ordered according to increasing powers of the participating variables, the most interesting dependence being one on the derivatives:

$b_0 = 1/4 \, \mu_0 \,( 1 + \mu_0 + 2\mu_1 )\,( 1 + 2\mu_0 + 2\mu_1 ),$

$D_0 = 47/32\mu_0 + 19/32\mu_1 + 121/32\mu_0^2 + 189/32\mu_0\mu_1 + 19/16\mu_1^2 + 2\mu_0^3 + 95/16\mu_0^2\mu_1 + 4\mu_0\mu_1^2 + ( -15/8\mu_0 - 63/16\mu_1 - 81/16\mu_0^2 - 123/8\mu_0\mu_1 - 45/4\mu_1^2 - 37/16\mu_0^3 - 163/16\mu_0^2\mu_1 - 29/2\mu_0\mu_1^2 - 27/4\mu_1^3)x_0 + ( -41/16\mu_0 - 3\mu_1 - 43/8\mu_0^2 - 13\mu_0\mu_1 - 123/16\mu_1^2 - 29/16\mu_0^3 - 53/8\mu_0^2\mu_1 - 131/16\mu_0\mu_1^2 - 27/8\mu_1^3)x_1 + (27/8\mu_0 + 81/16\mu_1 + 75/8\mu_0^2 + 387/16\mu_0\mu_1 + 243/16\mu_1^2 + 17/4\mu_0^3 + 37/2\mu_0^2\mu_1 + 399/16\mu_0\mu_1^2 + 81/8\mu_1^3)x_0^2 + (27/8\mu_0 + 81/16\mu_1 + 15/2\mu_0^2 + 171/8\mu_0\mu_1 + 243/16\mu_1^2 + 29/8\mu_0^3 + 241/16\mu_0^2\mu_1 + 339/16\mu_0\mu_1^2 + 81/8\mu_1^3)x_1^2$

$+(9/32 + 5/16\mu_0 + 1/8\mu_1 - 7/32\mu_0^2 - \mu_0\mu_1 - 7/8\mu_1^2 + ( -11/16 - 19/32\mu_0 - 9/32\mu_1 + 29/32\mu_0^2 + 91/32\mu_0\mu_1 + 35/16\mu_1^2)x_0 + ( -11/16 - 7/8\mu_0 - 1/2\mu_1 + 1/16\mu_0^2 + 3/2\mu_0\mu_1 + 7/4\mu_1^2)x_1 + (63/32 + 139/32\mu_0 + 159/32\mu_1 + 15/16\mu_0^2 + 103/32\mu_0\mu_1 + 33/16\mu_1^2)x_0^2 + (9/8 + 5/4\mu_0 + 15/16\mu_1 - 7/8\mu_0^2 - 51/16\mu_0\mu_1 - 21/8\mu_1^2)x_0 x_1 + (1/4 + 1/2\mu_0 + 1/2\mu_1 + 3/4\mu_0^2 + \mu_0\mu_1)x_1^2 + ( -27/16 - 75/16\mu_0 - 81/16\mu_1 - 17/8\mu_0^2 - 97/16\mu_0\mu_1 - 27/8\mu_1^2)x_0^3 + ( -27/16 - 15/4\mu_0 - 81/16\mu_1 - 29/16\mu_0^2 - 77/16\mu_0\mu_1 - 27/8\mu_1^2)x_0 x_1^2) \partial_0$

$+(1/16 + 7/16\mu_0 + 3/8\mu_1 + 5/8\mu_0^2 + 9/8\mu_0\mu_1 + 1/2\mu_1^2 + ( -1/2 - 2\mu_0 - 13/8\mu_1 - 31/16\mu_0^2 - 53/16\mu_0\mu_1 - 5/4\mu_1^2)x_0 + ( -1/2 - 25/16\mu_0 - 3/2\mu_1 - 17/16\mu_0^2 - 2\mu_0\mu_1 - \mu_1^2)x_1 + (1/2 + 15/8\mu_0 + 11/8\mu_1 + 25/16\mu_0^2 + 43/16\mu_0\mu_1 + 3/4\mu_1^2)x_0^2 + (15/16 + 51/16\mu_0 + 21/8\mu_1 + 19/8\mu_0^2 + 33/8\mu_0\mu_1 + 3/2\mu_1^2)x_0 x_1 + (2 + 69/16\mu_0 + 91/16\mu_1 + 23/16\mu_0^2 + 69/16\mu_0\mu_1 + 27/8\mu_1^2)x_1^2 + ( -27/16 - 75/16\mu_0 - 81/16\mu_1 - 17/8\mu_0^2 - 97/16\mu_0\mu_1 - 27/8\mu_1^2)x_0^2 x_1 + ( -27/16 - 15/4\mu_0 - 81/16\mu_1 - 29/16\mu_0^2 - 77/16\mu_0\mu_1 - 27/8\mu_1^2)x_1^3) \partial_1$

$+((9/32 +15/32\mu_0 +9/16\mu_1)x_0 +( -9/16 -15/16\mu_0 -9/8\mu_1)x_0^2 +( -11/16 -17/16\mu_0 -11/8\mu_1)x_0x_1 +(9/32 +15/32\mu_0 +9/16\mu_1)x_0^3$
$+(11/16 +17/16\mu_0 +11/8\mu_1)x_0^2x_1 +(1/4 +1/4\mu_0 +1/2\mu_1)x_0x_1^2) \partial_0^2$

$+(( -1/16 -1/8\mu_0 -1/8\mu_1)x_0 +(5/32 +7/32\mu_0 +5/16\mu_1)x_1 +(1/8 +1/4\mu_0 +1/4\mu_1)x_0^2 +( -1/16 -1/16\mu_0 -1/8\mu_1)x_0x_1 +( -1/32$
$+5/32\mu_0 -1/16\mu_1)x_1^2 +( -1/16 -1/8\mu_0 -1/8\mu_1)x_0^3 +( -3/32 -5/32\mu_0 -3/16\mu_1)x_0^2x_1 +( -1/8 -3/8\mu_0 -1/4\mu_1)x_1^3) \partial_0\partial_1$

$+(( -1/8 -1/4\mu_0 -1/4\mu_1)x_0 +( -1/16 -1/8\mu_0 -1/8\mu_1)x_1 +(7/16 +13/16\mu_0 +7/8\mu_1)x_0^2 +(1/8 +1/4\mu_0 +1/4\mu_1)x_0x_1 -1/16\mu_0x_1^2 +( -$
$3/8 -5/8\mu_0 -3/4\mu_1)x_0^3 +( -1/16 -1/8\mu_0 -1/8\mu_1)x_0^2x_1 +(1/16 +3/16\mu_0 +1/8\mu_1)x_1^3) \partial_1^2$.

The result for the other partial operator $D_1$ is somewhat more cumbersome (in accordance with the property iii), and both expressions shrink somewhat after substitution $\mu_i \to a_i + b_i\varepsilon$ with specific small integer $a_i$, $b_i$ within dimensional regularization, $\varepsilon = \tfrac{1}{2}(4-D)$. Note that constructing such $\varepsilon$-dependent answers directly for specific $a_i$, $b_i$ means one symbolic parameter less to deal with, and would result in a significant speedup which, however, would have to be balanced against the necessity to do separate calculations for different $a_i$, $b_i$.

In this regard one should keep in mind the first example of restoration of the analytical dependence on one of the kinematical parameters in the *D*-operators, the external momentum $k^2$ (see the project web-page [12]), where a direct calculation with a symbolic $k^2$ would have been prohibitively harder than several calculations for its specific values followed by an interpolation procedure. It is not completely clear how useful a systematic approach based on interpolations may eventually prove to be, but the first example is rather encouraging.

An important point to bear in mind is that all such answers — e.g. ones constructed by interpolations to restore symbolic dependencies — are easily verified by a direct substitution into eq. (2) within any general-purpose CAS. This protects one against major blunders given that one runs into degenerate cases once in a while in such calculations whose overall behavior is somewhat erratic.

**5 Conclusions**. The emergence of partial *D*-operators signifies a major theoretical simplification for the problem of generalized IBP reduction; it has allowed us to cross the two-loop barrier in this problem. As to the computing complexity, there are projects that far exceed what we have so far seen in our project (e.g. the $g-2$ effort [13], to name just one). Also, a number of planned programmatic optimizations are yet to be implemented, cf. [6].

On the mathematical side, there is evidence that our empirical algorithm does not yield simplest answers (remember that there is always a linear manifold of solutions to the Bernstein equation (3)). This is the case for the example with the result of ref. [5]; the two answers have the same complexity measured in terms of degrees with respect to $x$ and $\partial$, differing in the length of numerical coefficients. This implies that solutions like the one displayed above can be simplified (such simplifications are bound to be more significant for the more complicated examples). A heuristic recipe to achieve such a simplification would be desirable, not necessarily one that would yield an absolutely minimal solution because perfect solutions might be too expensive to construct and not required by the nature of the problem.

In any event, one can state (if only based on mere extrapolation of the number of surprises already uncovered in this problem) that further simplifications and optimizations are to be expected. For instance, an option to explore is the interpolation approach discussed at the end of sec. 4 as it has shown its potential with a successful restoration of the analytical dependence on a kinematical parameter where a direct algebraic approach is unrealistic. Numerical mathematics offers further algorithmic ideas for this analytical problem, and it would be nice to combine the benefits of the more abstract methods of ref. [5] with the brute-force approach of ref. [6].

**Acknowledgments**. We thank A. Czarnecki, G. Heinrich, M. Kalmykov and S. Volkov for stimulating interactions. This work was supported by R. Menyashev within the framework of the Informatika-21 project [14].